\documentclass[12pt]{article}

 \newcommand{\be}{\begin{equation}}
 \newcommand{\ee}{\end{equation}}
 \newcommand{\ba}{\begin{eqnarray}}
 \newcommand{\ea}{\end{eqnarray}}
 \newcommand{\bl}{\begin{equation}\begin{array}{ll}}
 \newcommand{\el}{\end{array}\end{equation}}
 \newcommand{\bll}{\begin{equation}\begin{array}{lll}}
 \newcommand{\bdm}{\begin{displaymath}}
 \newcommand{\edm}{\end{displaymath}}
 \def\p{\partial}
 \def\f{\varphi}
 \def\ve{\varepsilon}
 
\def\lim{\rightarrow}
\def\half{\frac{1}{2}}

\def\be{\begin{equation}}
\def\ee{\end{equation}}
\def\bea{\begin{eqnarray}}
\def\eea{\end{eqnarray}}

\begin{document}
\raggedbottom

\title{Integrable Models of Horizons and Cosmologies}

\author{A.T.Filippov\thanks{filippov@thsun1.jinr.ru}\\
{\small \it  Joint Institute for Nuclear Research, Dubna, Moscow region RU-142980}}


\maketitle

\begin{abstract}
A new class of integrable theories of 0+1 and 1+1 dimensional dilaton
gravity coupled to any number of scalar fields is introduced.
These models are reducible  to systems of independent Liouville
equations whose solutions must satisfy the energy and momentum constraints.
The constraints are solved thus giving the explicit analytic
solution of the theory in terms of arbitrary chiral fields.
In particular, these integrable theories describe spherically
symmetric black holes and branes of higher dimensional
supergravity theories as well as superstring motivated
cosmological models.
\end{abstract}

\section{Introduction}

In the last decade 1+1 and 0+1 dimensional dilaton gravity
coupled to scalar matter fields proved to be a reliable
model for higher dimensional black holes and string inspired
cosmologies. The connection between high and low dimensions has
been demonstrated in different contexts of gravity and string
theory - symmetry reduction, compactification, holographic
principle, AdS/CFT correspondence, duality. For spherically symmetric
configurations the description of static
black holes, branes, and of cosmological solutions
even simplifies to 0+1 dimensional dilaton gravity
- matter models, which in many interesting cases are explicitly
analytically integrable (see e.g. \cite{CGHS} - \cite{Iv}
and references therein).

However, generally they are not integrable. For example,
spherical black holes coupled to Abelian gauge fields are usually
described by integrable 0+1 dimensional models, while the addition of the
cosmological constant term destroys integrability. In 1+1 dimension,
pure dilaton gravity is integrable but the coupling to scalar matter
fields usually destroys integrability. The one very well studied exception
is the CGHS model. In \cite{A1} a more general
integrable model of dilaton gravity
coupled to matter, which incorporates as limiting
cases the CGHS and other known integrable models  was proposed.
It reduces to two Liouville
equations, whose solutions should satisfy two constraints.
Because the general analytic solution of the constraints had been
not found at that time, the model of Ref.\cite{A1} received little
attention and was not studied in detail\footnote{
Some applications of this model were discussed in \cite{Nav}. Note also that
some recent cosmological models use potentials similar to those introduced
in \cite{A1}. Such applications will be considered in a separate paper.}.

Recently, the author has proposed a class of more
general integrable dilaton gravity models in dimension 1+1, which
are reducible to $N$ Liouville equations (a brief summary is
published in Ref.\cite{A2}). For these models the general
analytic solution of the constraints has been found.
We demonstrated that the $N$-Liouville models are
closely related to physically interesting solutions of higher
dimensional supergravity theories describing the low energy limit of
superstring theories. These 1+1 dimensional $N$-Liouville theories describe
the solutions of higher dimensional theories in some approximation.
On the other hand,
their reduction to dimensions 1+0 (cosmological) and 0+1 (static
black holes) give the exact solution of higher dimensional theories.

Static black holes and cosmological models are described by one
dimensional solutions of the 1+1 dimensional theories. In the
standard approach the deep connection between black holes and cosmologies
 is not transparent and is usually ignored (even the precise
relation between the dimensional reductions used by `cosmologists' and
by `black holes investigators' is not quite obvious).
We thus start from the 1+1 dimensional formulation to get a unified
description of these two objects. A characteristic feature
of the static solutions of the models derived from string theory
is the existence of horizons with nontrivial scalar field
distributions (what must be characteristic features of string cosmologies
is as yet a much discussed problem).

It is well known that in the Einstein - Maxwell theories
minimally coupled to scalar fields the spherical static horizons disappear
if the scalar fields have a nontrivial space distribution (this is the so-called
 `no-hair theorem'). In Ref. \cite{A1}, a local version of the no-hair theorem
 (we call it the `no horizon' theorem) was formulated and proved. It states
that, under certain conditions, there exists no static solution with a
horizon in a class of 0+1 dimensional dilaton gravity
theories coupled to scalar matter (the important requirement
is that the scalar fields vary in space and are finite on the horizon).
The theorem is local, and does not require any boundary conditions at
infinity.

However, the `no horizon' theorem is not true (as is known in
several examples) for Einstein - Yang-Mills theories \cite{M1}
as well as for solutions of higher
dimensional supergravities, see e.g. \cite{Kiem}.
In all these cases the static solutions of
higher dimensional theories may be constructed by using the 1+1 or
0+1 dimensional dilaton gravity coupled to
matter. In the integrable models we discuss here the solutions
with horizons are completely identified and described in very
simple terms. One may also consider the global properties of the
solutions with or without horizons but we will not discuss this
subject here.

In Section~2 we briefly demonstrate that spherically
symmetric black holes and branes of higher dimensional
supergravity theories, as well as superstring motivated
cosmological models, may be described in terms of 0+1 and
1+1 dimensional dilaton gravity theories. In Sections~3,~4
a new class of integrable theories of 0+1 and 1+1 dimensional dilaton
gravity coupled to any number of scalar fields is introduced.
In Section~5 we discuss possible applications of the integrable
models and some unsolved problems.

\section{High dimensional dilaton gravity}

We first write the higher dimensional theories which, under
dimensional reductions, produce special examples of integrable
theories introduced in \cite{A1} and \cite{A2}. They all come from the low
energy limit of the superstring theories, which is described by
10 dimensional supergravities\footnote{
One may also start with the 11 dimensional supergravity, which is believed
to be related to the so called `M - theory', and reduce to 10 dimension
by compactifying one dimension. Note also that here we are not attempting to consider
compactifications of the most general supergravities. We only demonstrate
the main features of the connection between low dimensional and high dimensional
theories}.
The bosonic part of the 10
dimensional supergravities of type II (corresponding to the type
II superstrings) may be written as
 \be
 {\cal L}^{(10)} =
{\cal L}_{NS-NS}^{(10)} \, + \, {\cal L}_{RR}^{(10)} \, , \label{eq:1}
\ee
In this brief discussion it is sufficient to consider the first
Lagrangian\footnote{
The second one gives similar 1+1 dimensional theories.
Eventually, all bosonic and fermionic parts of the higher dimensional
Lagrangians give in 1+1 dimension dilaton gravity coupled to scalar matter
fields.}:
\be
{\cal L}_{NS-NS}^{(10)} \, = \sqrt{-g^{(10)}} e^{-2\phi_s} \biggl[ R^{(10)} +
4(\nabla \phi_s)^2 - {1\over 12} H_3^2\biggr] , \label{eq:2}
\ee
Here $\phi_s$ is the dilaton, related to the string coupling
constant; $H_3 = dB_2$ is a 3-form; $g^{(10)}$ and $R^{(10)}$ are the
10 dimensional metric and scalar curvature respectively.

There are many ways to reduce high dimensional theories to low dimensions
1+1 and 0+1. We only mention here those that may lead to integrable
theories. First, one may compactify the $d$ dimensional theory on
a $p$ dimensional torus $T^p$ (or on several tori, including the circle $S^1$)
using the Kaluza - Mandel - Fock - Klein (KMFK) mechanism\footnote{
It is usually called the Kaluza - Klein (KK) mechanism but this is not justified
historically. Actually, the Russian theorists George Mandel and Vladimir Fock have written
their papers, in which they generalized the Kaluza theory, even somewhat earlier
than Oscar Klein and published them in the same journal. We hope to redress this historical
injustice in a separate publication}.
This introduces $p$ Abelian gauge fields and at least $p$ scalar fields. Antisymmetric
tensor fields ($n$-forms), which may be present in the high dimensional theory,
will produce lower-rank forms and, eventually, other scalar fields. Thus we get
a theory of gravity coupled to matter fields (scalars, Abelian gauge fields and,
possibly, higher-rank forms) in a $d$ dimensional space, $d = D -p$. The next step
is to reduce further its dimension using some symmetry of the $d$ dimensional theory.
The most typical one is the spherical symmetry (the axial symmetry leads to much
more complex low dimensional theories and is not considered here). This step produces
a 1+1 dimensional dilaton gravity theory coupled to scalar and gauge fields.
The simplest example is the spherical reduction of the $d$ dimensional Einstein -
Maxwell theory - the resulting 1+1 dimensional dilaton gravity is actually equivalent
to a 0+1 dimensional theory.

The 1+1 dimensional dilaton gravity theories  so derived may describe static black holes
(static solutions), spherically symmetric evolution of the black holes (collapse of
matter) and of the universe (expansion of the universe). In this sense, the flat space
cosmological models and static black holes may be regarded as the 1+0 and 0+1
dimensional reductions of the 1+1 dimensional theory and they can be connected in the frame
of the 1+1 dimensional model.

Note that the final step in the chain of dimensional
reductions in cosmology is usually somewhat different from that in black hole physics.
Cosmological models are normally obtained by reducing the $d$ dimensional theory
directly to dimension 1+0. Indeed, isotropy and homogeneity of the universe require
a higher symmetry than the spherical one - the whole space should have constant
curvature $k$, which may be equal to zero or $\pm 1$. These cosmologies can be selected
from the set of the 1+0 dimensional solutions of the 1+1 dimensional theory by choosing
a proper dimensional reduction of the metric and of the dilaton. We will not go into
a detailed description of dimensional reductions, referring the reader to an instructive
example in \cite{BMG}, \cite{VDA}, \cite{Kiem} and to reviews \cite{St} - \cite{Iv}.
Instead we give a simplified typical chain of dimensional reductions leading to
simple and important two dimensional and one dimensional dilaton gravity models.

Reducing to $d$ dimensions by different sorts of dimensional
reduction (KMFK, compactification on tori, etc.) we obtain an
effective Lagrangian ${\cal L}^{(d)}$. For our purposes it is sufficient to consider
the following expression
\be
{\cal L}^{(d)} = \sqrt{-g^{(d)}} e^{-2\phi_d} \biggl( R^{(d)} + 4(\nabla
\phi_d)^2 - {1\over 2} (\nabla \psi)^2 - X_0 - X_1 (\nabla
\sigma)^2 - X_2 F_2^2 \biggr). \label{eq:3}
\ee
Here $\phi_d$ is a new dilaton, $F_2$ is a 2-form (an Abelian
gauge field); $X_a$ are functions of
$\phi_d$ and $\psi$. Actually, the Lagrangian should depend on
several $F_2$-fields, several $\psi$-fields, and may depend on several
$\sigma$-fields  as well as on higher - rank forms. However,
after further reduction to two dimension only 2-forms and scalar
fields will survive (in fact, the 2-forms can also be excluded by writing
an effective potential depending on electric or magnetic charges).

We further reduce the $d$ dimensional theory to dimension 1+1 by spherical
symmetry. Before and after doing so one may transform this Lagrangian by
the Weyl conformal transformation, $g_{\mu\nu} \Rightarrow \tilde{g}_{\mu\nu}
\equiv \Omega^2 g_{\mu\nu}$,
where $\Omega$ depends on the corresponding dilaton. Expressing $R$ in terms of the
new metric,
\be
R = \Omega^2 \biggl[\tilde{R} + 2(d-1) {\tilde{\nabla}}^2 \ln{\Omega} -
(d-1)(d-2) (\tilde{\nabla} \ln{\Omega})^2 \biggr] \, , \label{eq:3a}
\ee
one can easily find the new expression for the Lagrangian. For $d > 2$ one can
cancel the multiplier $e^{-2\phi_d}$ by choosing an appropriate function
$\Omega(\phi_d)$ and thus write the Lagrangian in the so called Einstein frame
(as distinct from the string frame expressions above). In dimension $d=2$ it is impossible
to remove the dilaton multiplier but, instead, one can remove the dilaton gradient
term.

Now consider the spherically symmetric solutions of the $d$ dimensional theory (3).
Usually, it is more convenient to remove the dilaton factor by
a Weyl transformation and rewrite the action
(3) in the Einstein frame,
\ba
{\cal L}^{(d)}_E = \sqrt{-g^{(d)}} \biggl[ R^{(d)} - \half (\nabla \chi)^2
- \half (\nabla \psi)^2 -
\nonumber \\
X_0 e^{a_0 \chi} - X_1 e^{a_1 \chi } (\nabla\sigma)^2 -
 X_2 e^{a_2 \chi } F_2^2 \biggr], \label{eq:3b}
\ea
where $\phi_d \equiv \chi$ and $a_k$ are known constants depending on $d$.
Then we parameterize the spherically symmetric metric by
the general 1+1 dimensional metric $g_{ij}$ and the dilaton $\f$ ($\nu \equiv 1/n$,
$n \equiv d-2$),
\be
ds^2=g_{ij}\, dx^i\, dx^j \,+\, {\f}^{2\nu} \,
d\Omega_{(d-2)}^2 \, , \label{eq:4}
\ee
introduce appropriate spherical symmetry conditions for the
fields, which from now on will be functions of the variables
$x_0$ and $x_1$ (or, $t$ and $r$), and integrate out the other (angular)
variables from the $d$ dimensional action.

Applying, in addition, the Weyl transformation that removes the dilaton gradient
term we thus obtain the effective 1+1 dimensional action
\ba
{\cal L}= \sqrt{-g} \biggl[ \f R + n(n-1) \f^{-\nu} -
X_0 e^{a_0 \chi } \f^{\nu} -
X_2 e^{a_2 \chi} \f^{2-\nu} F^2 -
\nonumber \\
 - \half \f \biggl( (\nabla \chi )^2
+ (\nabla \psi)^2 + 2 X_1 e^{a_1 \chi } (\nabla \sigma)^2
\,\biggr) \, \biggr] . \label{eq:5}
\ea
Here $\f$ is the 2-dilaton field that is often denoted by $e^{-2\phi}$;
the scalar fields $\psi$ may have different origins -- they may be former
dilaton fields, KMFK scalar fields, reduced $p$-forms, etc. The functions $X_k$ (we call
them potentials) depend on the scalar fields $\chi$ and $\psi$, which
from now on will be called scalar matter fields. Also the field $\sigma$ may be
regarded as a matter field but it plays a special role that will be discussed later.
Notice that the potentials are positive and that $n(n-1)$
is positive or zero\footnote{This term is the curvature of the
$n$ dimensional sphere whose metric is given by the second term in (5).
If, instead of the spherical symmetry,  we used
a pseudo-spherical one, the sign would be negative. If the $n$ dimensional
subspace is flat this term will be absent.}.

For dimensionally reduced supergravity theories one can often find a parameterization
of the fields in which the potentials are exponentials
of the matter fields while the kinetic (gradient) terms have the above simple structure.
These 1+1 dimensional theories may have integrable one dimensional
sector describing static (0+1) or cosmological (1+0) solutions of the higher dimensional
theories. The 1+1 dimensional theories obtained by dimensional reductions are
usually not integrable but may be approximated by explicitly analytically integrable
1+1 dimensional theories.

As it was mentioned above, the cosmological models are usually obtained from higher
dimensional theories by a different dimensional reduction. To describe the homogeneous
and isotropic universe one supposes that the metric may be written in the form
\be
ds^2 = -e^{2\nu(t)} dt^2 + e^{2\mu(t)} d\Omega^2_{(d-1)}(k) \, , \label{eq:5a}
\ee
where $k=0$ for the flat space and $k=\pm 1$ for the space of constant positive
(negative) curvature. Now, in cosmological models somewhat different reductions of
the fields are of interest because the terms generated by the higher rank forms
(characteristic of string theories) are believed to be of interest. Anyway, after
reducing to one dimension also the higher rank forms give scalar fields of either
$\psi$ or $\sigma$ type. For example, in a typical reduction of the type IIA
10 dimensional supergravity to dimension 4 (compactification on an isotropic
six dimensional torus $T^6$) and then to 1+0 dimensional dilaton gravity
(see e.g. \cite{Bill}), the 3-form produces in one dimensional theory a $\sigma$
term while the 4-form generates an $X_0$-type potential. The cosmologies  so obtained
are in general not integrable.


\section{1+1 dimensional dilaton gravity}

Now let us consider a general 1+1 dimensional dilaton gravity
coupled to Abelian gauge fields $F^{(a)}_{ij}$ and to scalar
fields $\psi_n$. The general Lagrangian can be written as
\bea
{\cal L} = \sqrt{-g} \, [ \, U(\f) R(g) +
V(\f) + W(\f)(\nabla \f)^2 + \nonumber \\
+ X (\f,\psi, F_{(1)}^2, ..., F_{(A)}^2) + Y(\f,\psi)+ \sum_n Z_n(\f,\psi)
(\nabla \psi_n)^2 \, ] \, . \label{eq:6}
\eea
Here $g_{ij}$ is a generic 1+1 dimensional metric with signature (-1,1) and
$R$ is the Ricci curvature; $U(\f)$, $V(\f)$, $W(\f)$ are
arbitrary functions of the dilaton field; $X$,
$Y$ and $Z_n$ are arbitrary functions of the
dilaton field and of $(N-2)$ scalar fields $\psi_n$ ($Z_n < 0)$;
$X$ also depends on $A$ Abelian gauge fields
 $F_{(a)ij} \equiv F^{(a)}_{ij}$,
 $F_{(a)}^2 \equiv g^{ii'} g^{jj'} F^{(a)}_{ij} F^{(a)}_{i'j'}$.
Notice that in dimensionally reduced theories (see (\ref{eq:5}))
both the scalar fields and the Abelian gauge fields are
non-minimally coupled to the dilaton.

The equations of the theory (\ref{eq:6})
can be solved for arbitrary potentials $U$, $V$, $W$ and $X$ if
$\p_{\psi}X \equiv 0$
(for the simplest explicit solution in case of $X$ linear in $F^2$
see e.g \cite{A1} and references therein as well as the recent review \cite{Kummer}).
Actually, only $V(\phi)$ and $X$ are really arbitrary functions.
Moreover, for general potentials $X(\phi, \psi, F^2)$ one may solve the equations for
$F^{(a)}_{ij}$ and then construct the effective action
\be
{\cal L}_{\rm eff} =\sqrt{-g}\left[ \f R(g) + V_{\rm eff}(\f,\psi) +
\sum_n Z_n(\f,\psi)\, g^{ij} {\p}_i \psi_n {\p}_j \psi_n \right]\,.
\label{eq:7}
\ee
Here the effective potential $V_{\rm eff}$ (below we omit the subscript) depends
also on charges introduced by solving the equations for the Abelian
fields. Note also that we use a Weyl transformation to exclude the kinetic
term for the dilaton and also choose the simplest, linear parameterization
for $U(\f)$\footnote{If $U^{\prime}(\f)$ has zeroes, this parameterization, as well as
the more popular exponential one, $U = e^{\lambda \phi}$, is valid only between
two consecutive zeroes.}.

If  the effective potential does not depend on $\psi$, one can find the general
solution for the matter vacuum when all $\psi$ are constant. In this case the
equation of motion actually reduce to those of the pure dilaton gravity not
coupled to scalar matter. Few 1+1 dimensional models are integrable. The best studied ones
are the CGHS and JT models. They were essentially generalized in \cite{A1}.
In all these models $Z$-potentials are constant (so called minimal coupling).
The only integrable class of models with non minimal coupling to scalar fields
(with some special functions $Z_n(\phi)$) was proposed in \cite{FI}.

For the sake of completeness we write here
the equations of motion in the light cone ($u,v$) coordinates
(see e.g. \cite{A1}; to simplify the formulas we keep only one scalar field
and omit the superscript of the effective potential):
\ba
&& \p_u \p_v \f+f\, V(\f,\psi)=0, \label{F.15} \\
&&\p_v (Z \p_u \psi) +\p_u (Z \p_v \psi) + f V_{\psi}(\f,\psi)=
Z_{\psi} \, \p_u \psi \, \p_v \psi,
\label{F.16} \\
&& f \p_i ({ {\p_i \f} \over f }) \, = Z (\p_i \psi)^2, \quad i=u,v, \label{F.17} \\
&&\p_u\p_v\ln |f| + f V_{\f}(\f,\psi) = Z_{\f} \,\p_u \psi \, \p_v\psi,
\label{F.18}
 \ea
 where $V_{\f}= \p_{\f} V$, $V_{\psi}= \p_{\psi} V$,
 $Z_{\f}= \p_{\f} Z$, $Z_{\psi} =\p_{\psi} Z$.
  Equations (\ref{F.15}) $-$ (\ref{F.18}) are not independent. Actually,
  (\ref{F.18}) follows from equations
(\ref{F.15}) $-$  (\ref{F.17}). Alternatively, if  (\ref{F.15}),
(\ref{F.17}), (\ref{F.18}) are satisfied,
   (\ref{F.16}) is satisfied.

Now we introduce a more general class of integrable
1+1 dimensional dilaton gravity models with minimal coupling to scalar fields.
They are defined by the Lagrangian (\ref{eq:7})
with the following potentials:
\be
Z_n = -1; \;\;\;\;  |f|V = \sum_{n=1}^N 2g_n e^{q_n} \, . \, \label{eq:8}
\ee
Here $f$ is the light cone metric, $ds^2 = -4f(u, v) \, du \, dv$, and
\be
q_n \equiv F + a_n \phi + \sum_{m=3}^{N} \psi_m a_{mn} \equiv
\sum_{m=1}^{N} \psi_m a_{mn} \, , \label{eq:9}
\ee
where $\psi_1 + \psi_2 \equiv \ln{|f|} \equiv F$ ($f \equiv \varepsilon e^F$)
and $\psi_1 - \psi_2 \equiv \phi$.

Now, varying the Lagrangian (\ref{eq:7}) in $(N-2)$ scalar fields, dilaton and in
$g_{ij}$ and then passing to the light cone metric we find $N$ equations of motion
for $N$ functions $\psi_n$ ,
\be
\epsilon_n \p_u \p_v \psi_{n} =  \sum_{m=1}^{N} \varepsilon g_me^{q_m} a_{mn} \, ; \,\,\,\,
\epsilon_1 = -1, \,\,\, \epsilon_n = +1, \,\, {\rm if} \,\, n \geq 2 \, ,
\label{eq:10}
\ee
and two constraints,
\be
C_i \equiv f {\p}_i ({\p}_i \phi /f) + \sum_{n=3}^N ({\p}_i \psi_n)^2 = 0 ,
\,\,\, i = (u, v) .
\,\, \label{eq:11}
\ee


For arbitrary coefficients $a_{mn}$ the equation of motion are not integrable.
However, if the $N$-component vectors $v_n \equiv (a_{mn})$ are
pseudo-orthogonal, the equations of
motion can be reduced to $N$ Liouville equations for $q_n$,
\be
{\p}_u {\p}_v q_n - {\tilde{g}}_n e^{q_n} =0 , \label{eq:12}
\ee
where ${\tilde{g}}_n = \ve \lambda_n g_n$, $\lambda_n = \sum \epsilon_m a_{mn}^2$,
and $\ve \equiv |f|/f$ (note that the equations for $q_n$ depend on $\epsilon_n$
only implicitly, through the normalization factor $\lambda_n$).

The most important fact is that the constraints can be explicitly solved.
By writing the solutions of the Liouville equations in the form suggested
by the conformal field theory,
\be
e^{-q_n /2} = a_n(u) b_n(v) + \bar{a}_n(u) \bar{b}_n(v) ,
\label{eq:13}
\ee
where $\bar{a}$ and $\bar{b}$ can be expressed in terms of $a$ and $b$, i.e.
\be
e^{-q_n /2} = a_n(u) b_n(v) \biggl[ 1 - \half {\tilde{g}}_n
\int {du \over a_n^2(u)} \int {du \over b_n^2(v)} \biggr] \, ,
\ee
we may rewrite the constraints in the form
\be
C_u = \sum_{n=1}^N a_n^{\prime \prime} (u) [\lambda_n a_n(u)]^{-1} \, , \,\,\,\,
C_v = \sum_{n=1}^N b_n^{\prime \prime} (v) [\lambda_n b_n(v)]^{-1} \, .
\label{eq:14}
\ee
Using the fact that the norms $\lambda_n$ satisfy the constraint
$\sum_{n=1}^N \lambda_n^{-1} = 0$ (this is a consequence of the pseudo-orthogonality
conditions) we can solve these constraints. The solution has the
following form:
\be
{{a_n^{\prime} (u)} \over {a_n(u)}} = \alpha_n (u) -  {{\sum_{n=2}^N {\lambda_n}^{-1}
(\alpha_n^{\prime} + \alpha_n^2)} \over {2\sum_{n=2}^N {\lambda_n}^{-1} \alpha_n}} ,
\label{eq:15}
\ee
where $\alpha_1 = 0$ and the other $\alpha_n$ are arbitrary functions of $u$.
The ratios $b_n^{\prime}(v) / b_n(v)$ are expressed by the same equation in terms
of functions $\beta_n(v)$.

By integrating the first order differential equations for $a_n(u)$ and $b_n(v)$
we thus find the general solution of the $N$-Liouville dilaton gravity in terms
of $(2N-2)$ arbitrary chiral fields $\alpha_n(u)$ and $\beta_n(v)$.
With proper conditions of convergence one may use this solution also for $N=\infty$.


\section{Integrable 0+1 dimensional dilaton gravity coupled to matter}

The dimensional reduction from 1+1 to 0+1 dimensions in the light cone
coordinates $(u, v)$ is very simple.
 If we suppose that $\f=\f(\tau)$, $\psi_n =\psi_n(\tau)$ where
 $\tau=a(u)+b(v)$, we find from the 1+1 dimensional equations of
 motion that
 \be
 f(u,v) = \mp h(\tau) \, a'(u) b'(v) ; \,\,\,\,\,
 ds^2 = -4f \, du \, dv  = \pm 4h(\tau) da db. \,\,
 \label{eq:16}
 \ee
Defining two dimensional space and time coordinates, $r = a \pm b$ and
$t = a \mp b$ we find that
\be
ds^2 = h(\tau) (dt^2 - dr^2) , \,\,\,\, {\rm where} \,\,\,\,
\tau = r \,\, {\rm or} \,\, t ,
\label{eq:17}
\ee
and thus the reduced solution may be either static or cosmological
one\footnote{Of course, in 2d theories this distinction is not very important.
However, when we know the higher dimensional theory from which our 2d dilaton
gravity originated, we can reconstruct the higher dimensional metric and thus
find the higher dimensional interpretation of our solutions. In the remaining part
of this paper we do not use $r$ and $t$, take in (\ref{eq:16}) the upper sign
and usually call all one dimensional solutions static.}.

However, this is not the most general way for obtaining 0+1 or 1+0 dimensional
theories from higher dimensional ones. Not all possible reductions can be derived
by this simple dimensional reduction of the 1+1 dimensional gravity. For example,
to derive the cosmological solutions corresponding to the reductions (\ref{eq:5a})
one should use a more complex dimensional reduction of the 1+1 dimensional dilaton
gravity, which will be discussed elsewhere.

It is not difficult to show that the 0+1 dimensional equations are described
by the Lagrangian ($F \equiv \ln{|h|}$, $\varepsilon \equiv \pm$) \cite{A1}:
\be
{\cal L} =-{1\over l} \biggl( \dot \f \dot F +\sum_n
Z_n(\f, \psi) \dot \psi_n^2 \biggr) +l\, \ve e^F \, V(\f , \psi) ,
\label{eq:18}
\ee
where $l(t)$ is the Lagrange multiplier (related to the general metric $g_{ij}$).

Now, the two - dimensionally integrable $N$-Liouville theories presented above are also
integrable in 0+1 dimension. Moreover, as we can solve the Cauchy problem in dimension
1+1 we can study the evolution of the initial configurations to stable static solutions,
e.g. black holes, which are special solutions of the 0+1 dimensional reduction.
However, the reduced theories can be explicitly solved for much more general potentials
$Z_n$ and $V$.

Suppose that for $(N-2)$ scalar fields $\psi_n$ ($n = 3,...,N$) the
ratios of the $Z$-potentials are constant so that we can write
$Z_n = -\gamma_n /\phi^{\prime}(\f)$ (in eq.~(\ref{eq:5}) these are the fields $\chi$ and
$\psi$ and $\phi = \ln \f$). Suppose that all the potentials $Z_n$ and $V$ be
independent of the scalar fields $\psi_{N+k}$ with $k = 1,...,K$ (in eq.~(\ref{eq:5})
this is the field $\sigma$). Then, we first remove the factor $\phi^{\prime}(\f)$
by defining the new Lagrange multiplier $\bar{l} = l(\tau) \phi^{\prime}(\f)$ and absorb the
factors $\gamma_n > 0$ in the corresponding scalar fields. In this way we may introduce
the new dynamical variable $\phi$ instead of $\f$. Now we can solve the equations for
the $\sigma$-fields and construct the effective Lagrangian\footnote{
It is better to do this in the Hamiltonian formalism but space limitations force us
to omit details of our derivations.}.
We thus may arrive at the effective Lagrangian
\be
{\cal L}_{\rm eff} =-{1\over \bar{l} }\biggl[ \dot\phi \dot F - \sum_{n=3}^N
 \dot \psi_n^2 \biggr] + \bar{l} \biggl[ \ve e^F  V_{\rm eff}(\phi, \psi) +
 V_{\sigma} (\phi, \psi) \biggr] .
\label{eq:19}
\ee
Here $V_{\rm eff} = V/\phi^{\prime}(\f)$ and $V_{\sigma} =
\sum_k C_k^2 /Z_{N+k} \, \phi^{\prime}(\f)$ where $\f$ must be expressed in terms of
$\phi$.
If the original potentials in eq.~(\ref{eq:18}) are such that
$(Z_{N+k} \, \phi^{\prime}(\f))^{-1}$ and
$V / \phi^{\prime}$ can be expressed in terms of sums of exponentials of linear combinations
of the fields $\phi$ and $\psi$, then there is a chance that the 0+1 dimensional
theory can be reduced to Liouville or Toda equations (the Toda case is possible
only if $V_{\sigma} \neq 0$).

The pure Liouville case was introduced in \cite{A2}
and is described by the Lagrangian (in notation of eq.~(\ref{eq:9}))
\be
{\cal L} = {1\over l} \bigl( - \dot \psi_1^2 +
 \sum_{n=2}^N \dot \psi_n^2 \bigr) +l \sum_{n=1}^N 2g_n
e^{q_n}.
\label{eq:20}
\ee
If the $a_{mn}$ satisfy our pseudo-orthogonality conditions, the equations of motion are
reduced to $N$ independent one dimensional Liouville equations whose solutions
have to satisfy the energy constraint. The solutions are expressed in terms of
elementary exponentials (for simplicity, we write the solution in the gauge
$l(\tau) \equiv 1$ but all the results are actually gauge invariant):
\be
e^{-q_n}={|\tilde{g}_n|\over 2\mu^2_n} \biggl[ e^{\mu_n (\tau-\tau_n)}
+ e^{-\mu_n(\tau-\tau_n)} +2\ve_n \biggr] ,
\label{eq:21}
\ee
where $\ve_n \equiv -|\tilde{g}_n| /\tilde{g}_n$, $\mu_n$ and $\tau_n$ are
the integration constants ($\mu_n^2$ and $\tau_n$ are real).
The constraint can be shown to be $\sum_n \mu_n^2 /\lambda_n = 0$,
and its solution is trivial. The space of the solutions is thus defined by the
$(2N-2)$ dimensional module space (one of the $\tau_n$ may be fixed).
One can show that the solutions have at most 2 horizons\footnote{
To prove this one should analyze the behavior of $q_n$ for $|\tau| \lim \infty$
and for $|\tau - \tau_n | \lim 0$ (if $\ve_n < 0$). The horizons appear when
$F \lim -\infty$ while $\phi$ and $\psi_n$ for $n \geq 2$ tend to finite limits.
This is possible for $|\tau | \lim \infty$ if and only if $\mu_n = \mu$.
When $F \rightarrow F_0$ and $\phi \rightarrow \infty$ we have the flat space limit,
e.g. exterior of the black hole.
The singularities in general appear for $|\tau - \tau_n | \lim 0$ if $\ve_n < 0$.}
and the space
of the solutions with horizons has dimension $(N-1)$. There exist integrable models
having solutions with horizons and no singularities but their relation to the high
dimensional world is at the moment not clear.

Note that the solution (\ref{eq:21}) is written in a rather unusual coordinated.
One may write a more standard representation remembering that the dilaton $\phi$
is related to the coordinate $r$ (see \ref{eq:4})). This may be useful for a geometric
analysis of some simple solutions (e.g. Schwarzschild or Reissner - Nordstr\o m) but in general
the standard representation is very inconvenient for analyzing solutions of the
$N$-Liouville theory.

\section{Discussion and outlook}
The explicitly analytically integrable models presented here
may be of interest for different applications.
The most obvious is to use them to construct first approximations to generally
non integrable theories describing black holes and cosmologies. Realistic theories
of these object are usually not integrable (even in dimension 0+1). Having explicit
general solutions of the zeroth approximation in terms of elementary functions
it is not difficult to construct different sorts of (classical) perturbation theory.

For example, spherically symmetric static black holes non minimally coupled to
scalar fields are described by the integrable 0+1 dimensional $N$-Liouville
model. However, the corresponding 1+1 dimensional theory is not integrable
because the scalar coupling potentials $Z_n$ are not constant
(see eq.(\ref{eq:5})). To obtain approximate analytic solutions of the 1+1
dimensional theory one may try to approximate $Z_n$ by properly chosen constants.

It may be useful to combine this approach with the recently proposed analytic perturbation
theory allowing to find solutions near horizons
for most general non integrable 0+1 dilaton gravity theories \cite{atfm}.
The detailed description of the $N$-Liouville (and of the Toda type) theory
as well as applications to black holes and cosmology will be given elsewhere \cite{DAF}.
The Toda type theories were earlier introduced by direct reductions of higher dimensional
theories to cosmological models
(see e.g.  \cite{Iv}, \cite{Fre} and references therein).

The author is grateful to V.~de Alfaro for very useful collaboration and warm hospitality
at the University of Turin, to P.~Fre, O.~Lechtenfeld, D.~Luest, D.~Maison and A.~Sorin
for interest in this work and useful remarks, and to D.~Luest for hospitality at the Humboldt
University in Berlin. This work was partly supported by RFBR grant 03-01-00781.\footnote{
This paper is an extended version of the report that will be published in the Proceedings
of the third Sakharov Conference (sc3). Neither the report nor earlier papers \cite{A2}
in which I very briefly described the integrable models discussed here were sent to hep-th.
A more detailed presentation of the models and their applications will be given \cite{DAF}.}

\section{Appendix}

To help the reader in keeping trace of relations between dimensions 1+1, 1+0 and 0+1 we
write here a simple expression for the curvature in dimension 1+1. We take the diagonal
metric
\bdm
ds^2 = -e^{2\nu}dt^2 + e^{2\mu}dr^2 .
\edm
The Ricci scalar $R$ for this metric is
\bdm
R = 2 e^{-2\nu} (\ddot{\mu} + {\dot{\mu}}^2 - \dot{\mu} \dot{\nu} ) -
2 e^{-2\mu} (\mu^{\prime\prime} + {\nu^{\prime}}^2 - \nu^{\prime} \mu^{\prime}) .
\edm
For this metric, one may also need the expression for $\nabla^2 \phi$,
where $\phi$ is an arbitrary scalar field:
\bdm
\nabla^2 \equiv \nabla^m \nabla_m \phi = - e^{-2\nu} \bigl( \ddot{\phi} +(\dot{\mu} -
\dot{\nu}) \dot{\phi}) + e^{-2\mu} ({\phi}^{\prime\prime} +
({\nu}^{\prime} - {\mu}^{\prime}) {\phi}^{\prime} \bigr) .
\label{a6}
\edm
In the $(u,v)$ coordinates the curvature $R$ is simply
\bdm
R = {1 \over f} \p_u \p_v \ln{|f|} .
\edm
Note that the equations of motion (but not the constraints) may be
derived from the gauge fixed Lagrangian
\bdm
{\cal L}=\f \, \p_u\p_v F + f V-Z \, \p_u \psi \, \p_v\psi .
\edm
It is easy to derive from it the dimensionally reduced one dimensional
equations of motion and the Lagrangian (\ref{eq:18}) restoring the constraint,
if one passes to the Hamiltonian formalism and recalls that the constraint
in one dimension requires that the Hamiltonian should vanish (see \cite{A1}).

\end{document}